\PassOptionsToPackage{table}{xcolor}
\documentclass[sigconf,screen]{acmart}

\AtBeginDocument{%
  \providecommand\BibTeX{{%
    \normalfont B\kern-0.5em{\scshape i\kern-0.25em b}\kern-0.8em\TeX}}}

\copyrightyear{2026}
\acmYear{2026}
\setcopyright{cc}
\setcctype{by}
\acmConference[DIS '26]{Designing Interactive Systems Conference}{June 13--17, 2026}{Singapore, Singapore}
\acmBooktitle{Designing Interactive Systems Conference (DIS '26), June 13--17, 2026, Singapore, Singapore}
\acmDOI{10.1145/3800645.3812922}
\acmISBN{979-8-4007-2563-0/2026/06}

\usepackage[T1]{fontenc}
\usepackage[utf8]{inputenc}

\usepackage{booktabs}
\usepackage{dcolumn}
\usepackage{tabularx}
\usepackage{framed}
\usepackage{multirow}
\usepackage{enumitem}
\usepackage{lscape}
\usepackage{longtable}
\usepackage{siunitx}
\usepackage{microtype}
\usepackage{etoolbox}
\usepackage{makecell}
\usepackage{subcaption}
\usepackage{multicol}
\usepackage{tikz}
\usepackage{array}

\newcolumntype{M}[1]{>{\centering\arraybackslash}m{#1}}

\begin{document}
\author{Sujay Shalawadi}
\email{sujay.shalawadi@ntnu.no}
\orcid{0000-0003-3937-5427}
\affiliation{%
  \department{Department of Design}
  \institution{Norwegian University of Science and Technology}
  \city{Gjøvik}
  \country{Norway}
}

\author{Joel Wester}
\orcid{0000-0001-6332-9493}
\email{joel.wester@di.ku.dk}
\affiliation{%
  \department{Department of Computer Science}
  \institution{University of Copenhagen}
  \city{Copenhagen}
  \country{Denmark}
}

\author{Samuel Rhys Cox}
\email{srcox@cs.aau.dk}
\orcid{0000-0002-4558-6610}
\affiliation{%
  \department{Department of Computer Science}
  \institution{Aalborg University}
  \city{Aalborg}
  \country{Denmark}
}

\author{Niels van Berkel}
\email{nielsvanberkel@cs.aau.dk}
\orcid{0000-0001-5106-7692}
\affiliation{%
  \department{Department of Computer Science}
  \institution{Aalborg University}
  \city{Aalborg}
  \country{Denmark}
}

\title[User Tensions with AI-Generated Fitness Feedback]{Who Gets to Interpret the Workout? User Tensions with AI-Generated Fitness Feedback}

\renewcommand{\shortauthors}{Shalawadi et al.}

\begin{abstract}
Fitness tracking platforms increasingly integrate generative AI to interpret activity data, such as Strava's Athlete Intelligence. These integrations raise questions about how athletes engage with AI-supported fitness self-tracking. We analyzed 297 Reddit threads and 5,692 comments from \texttt{r/Strava} following the company's launch of AI features to examine user reactions to AI-generated fitness feedback. Our findings revealed four recurring tensions: (1) numerical evaluation versus contextual understanding; (2) isolated session summaries versus ongoing training narratives; (3) a fixed AI tone versus diverse emotional states; and (4) a single AI voice versus different athletic types. Across these tensions, users resisted AI feedback that constrained interpretations of their own lived experiences. These findings shed light on the implicit challenges of integrating AI into self-tracking platforms. We conclude with implications for the design of AI-supported self-tracking systems that preserve interpretive openness and user agency.
\end{abstract}

\begin{teaserfigure}
  \includegraphics[width=\textwidth]{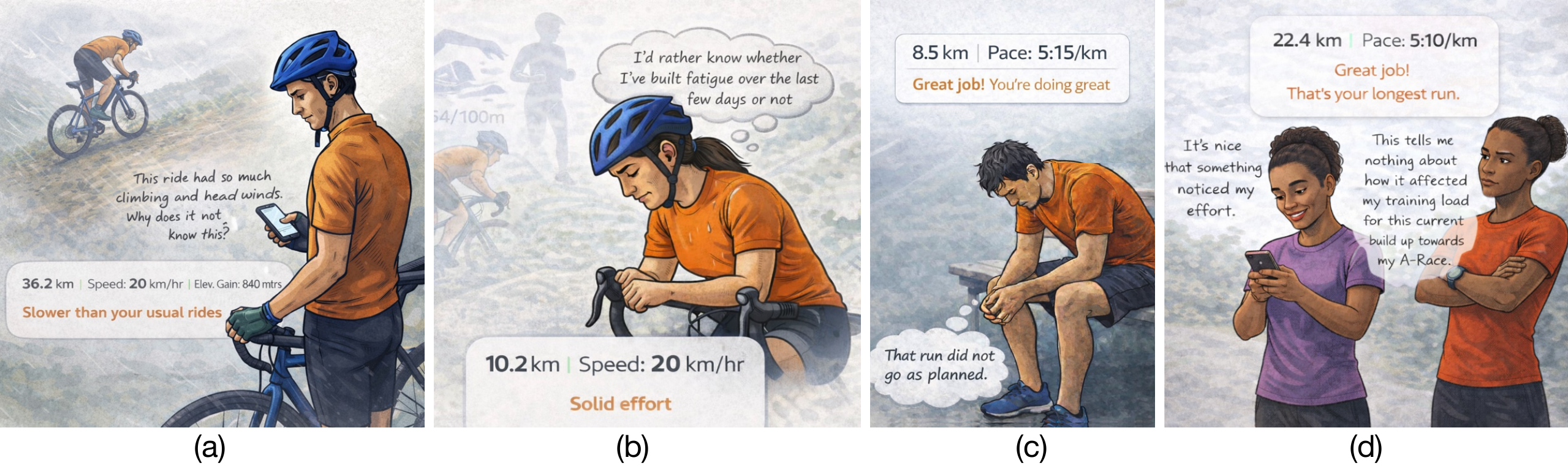}
  \caption{The four tensions identified in athletes' experiences with Athlete Intelligence: (a) AI's Numerical View versus Athletes' Contextual View; (b) Isolated Summaries versus an Ongoing Training Narrative; (c) a fixed, upbeat AI tone inappropriate for diverse emotional states; and (d) Uniform AI Voice versus different Athlete types.}
  \Description{A teaser figure summarizing four tensions in athletes' experiences with AI-generated fitness feedback.}
  \label{fig:header}
\end{teaserfigure}



\maketitle

\section{Introduction}

Strava is a widely used digital platform for tracking fitness activities, with over 180 million registered users across 190 countries~\cite{strava2024gift,StravaUserProfile}. The platform enables individuals to log and review their workouts alongside social features such as commenting on each other's activities and competing on leaderboards~\cite{Couture02012021,russell2023if,hochstetler2024kudos,RiversStrava}. Through activity logging and social sharing, fitness-tracking platforms such as Strava can support accountability~\cite{epstein2020exploring,wang2024exploring}, sustain motivation~\cite{Chung}, and foster recognition and encouragement within athletic communities~\cite{Kunwoo,Gui}.
A central part of these practices is the work athletes do to make sense of their activity data. Training is interpreted through accumulated experience, bodily sensations, environmental conditions, emotional states, and longer-term goals~\cite{epstein2015livedinformatics,Loerakker2025lived,Niess_Rumination}. Interpretation is therefore not a single analytic step, but an ongoing and situated practice that unfolds across workouts and over time. 

Recently, fitness platforms have begun experimenting with automating parts of this practice. In October 2024, Strava introduced Athlete Intelligence, a generative AI feature that automatically produces narrative summaries of users' activities~\cite{Athlete_Intelligence}. 
This feature is marketed as offering athletes personalized insights by aggregating recent activity data into a textual account attached to each workout. 
In personal informatics (PI) research, this process is termed `integration': the combination and transformation of data for analysis~\cite{Moore}.
Integration is distinct from reflection, which users carry out for themselves as they interpret what data means for their goals and circumstances~\cite{rapp2017knowthyself}.
By generating narrative summaries automatically, Athlete Intelligence performs integration on the user's behalf and presents the results as an interpretation of the workout. 
Introducing generative AI at this stage, where athletes have traditionally drawn on situated experience and personal judgment~\cite{rapp2017knowthyself,Moore}, raises questions about how such feedback shapes self-tracking practices.

In this paper, we examine how athletes interpret and contest AI-generated summaries. We do so by analyzing discussions from 297 threads and 5,692 comments by 2,611 unique contributors on the subreddit \texttt{r/Strava} following the launch of Athlete Intelligence between October 2024 and March 2025. Our qualitative analysis shows that users did not treat AI-generated summaries as definitive explanations of their workouts. Instead, they compared these summaries with their own training knowledge, bodily experience, situational context, and longer-term goals. We identified four recurring types of tensions, as illustrated in~\autoref{fig:header}. While some users appreciated the convenience or encouragement the summaries offered, most found them misaligned with their lived training experiences.

Building on these findings, this paper contributes to research on self-tracking technologies in three ways. First, we provide an empirical account of how athletes engage with AI-generated feedback that is automatically attached to their activity records, identifying four recurring tensions between AI-generated interpretations and users’ situated understandings of their training. Second, we discuss how these tensions arise from differences between how AI-generated summaries interpret workouts and how athletes understand their own training, often leading to competing interpretations. Third, we derive design implications for AI-first integration in self-tracking systems, highlighting how such systems can support interpretation without constraining it.

\section{Related Work}

We begin by contextualizing sense-making in the integration phase of self-tracking. Next, we review prior work on AI-generated integration in fitness technologies. We then examine how platform-mediated interpretations shape meaning and user agency, before summarizing the research gap this paper addresses.

\subsection{Integration and Sense-Making in Self-Tracking}

Self-tracking technologies support people in turning raw data into interpretations that make activities legible and meaningful~\cite{li2010personalinformatics,epstein2015livedinformatics}. Early models framed this process as a clean progression through predefined stages. Subsequent research, however, shows that sense-making is deeply entangled with everyday life shaped by routines~\cite{Rainmaker} rather than unfolding as a stable analytic sequence. Rooksby et al.~\cite{rooksby2014livedinformatics} demonstrate that people track in episodic and situated ways shaped by routine and frequent lapses. Niess and Wozniak~\cite{niess2018trackergoal} show that interpretation is closely tied to shifting goals and motivation, with users making sense of data in response to progress, frustration, or disengagement. Loerakker et al.~\cite{Loerakker2025lived} further highlight the role of embodied experience, showing that understanding self-tracked data is inseparable from how the body feels during and after activity. This perspective is reinforced by \cite{Tholander}, who demonstrates that athletes often rely on bodily sensations as a primary source of knowledge when interpreting performance, rather than treating metrics as definitive representations. Together, this work positions sense-making as a lived and dynamic practice rather than a discrete analytic step.

Within this broader process, personal informatics research distinguishes integration, the computational transformation of collected data into summaries or insights, from reflection, which users carry out for themselves~\cite{epstein2015livedinformatics}. Interpretation unfolds over time and is influenced by emotional context, physical state, and longer-term goals~\cite{rapp2020selftrackingsport,rapp2023livedexptechnologies}. At the same time, platforms increasingly perform integration by detecting trends, surfacing metrics, or generating narrative summaries. These platform-authored interpretations can shape how athletes understand their progress and identity~\cite{rapp2017knowthyself,eikey2021beyondreflection}, particularly when they foreground certain readings of an activity over others.

Despite its importance, integration remains one of the least examined stages in PI. A review of more than 500 studies shows that HCI research has focused primarily on supporting data collection and reflection, devoting far less attention to the analytic processes that connect them~\cite{epstein2020mappingpersonalinformatics}. Moore et al.~\cite{Moore} describe this imbalance as the personal informatics analysis gap, noting that many systems rely on rigid, designer-defined analytic pipelines that do not reflect how people actually want to interrogate or combine their data. 
Recent AI features have drawn renewed attention to this gap by enabling platforms to perform integration on behalf of users at scale. Understanding how people respond to these automated interpretations is therefore critical for examining how meaning is constructed in contemporary self-tracking systems.

\subsection{AI-Generated Integration in Fitness Tracking Technologies}

Fitness tracking technologies increasingly use AI to generate summaries or produce narrative interpretations of activity data. 
Recent systems provide textual accounts of workouts, training load, or progress in order to simplify complex metrics and offer quick, high-level assessments~\cite{NarratingFitness,TrainingLoad,MoveAI}.
Most research on such systems examines tools that users actively choose to engage with.
For example, prior work examines how individuals respond to such AI-generated interpretations. Duking et al.~\cite{duking2024chatgpt} report that conversational AI tools can provide helpful structure, yet their advice often overlooks environmental or bodily factors that users consider central to understanding performance. Namvarpour et al.~\cite{Namvarpour}, studying interactions with relationship-oriented AI agents, show that users frequently reject motivational language when it appears generic or insensitive to emotional context. Although outside the fitness domain, their findings highlight a broader challenge for AI-generated narratives, namely that users are skeptical of interpretations that do not align with lived experience or intent.

Several studies focus specifically on AI-based fitness and health assistants. GPTCoach examines an LLM-based fitness assistant and finds that users value motivational structure but often perceive its narrative interpretations as impersonal or misaligned, particularly when the system overlooks effort, fatigue, injury, or training intent~\cite{GPTCoach}. PhysioLLM presents an interactive system that integrates wearable physiological data with contextual information to generate personalized health interpretations. Its evaluation shows that domain-aware, data-grounded AI interpretations support a deeper understanding of personal data than generic LLM outputs~\cite{PhysioLLM}. Together, these studies underscore the importance of contextual and analytic specificity in AI-generated integration.

Related work also points to variation in how users want interpretations to be presented~\cite{Tunis}. Some prefer concise summaries that foreground salient aspects of a workout, while others seek richer explanations or more granular breakdowns~\cite{bentvelzen2023fitnesstrackers,mitchell2021nutritiongoal}. Studies on explainability and adaptive feedback further suggest that transparency into how interpretations are produced affects whether users perceive them as credible and personally meaningful~\cite{kocielnik2018reflectioncompanion,GoalsforGoal}.

Overall, prior work shows that people evaluate AI-generated interpretations in relation to context, bodily experience, goals, and autonomy. 
However, the systems examined in this literature share a common feature: users initiate engagement with the AI. 
Far less is known about how people respond when a platform generates an interpretation automatically and attaches it to an activity record by default.
In such cases, users encounter the AI's framing before reflecting on the activity themselves, potentially altering the dynamics observed in studies of requested feedback.

\subsection{Platform Mediation of Interpretation}

Self-tracking platforms do more than record activity data; they increasingly mediate how that data is interpreted. Features such as automatic captions, inferred metrics, or highlighted trends allow platforms to generate representations of workouts before users provide their own accounts~\cite{Couture02012021}. This shift raises questions about authorship and agency, as platform-generated interpretations can influence how individuals understand their activity and how it is presented~\cite{GenAI}.

Research on algorithmic agents and automated interventions in other domains offers relevant perspectives on this form of mediation. A common finding is that automated interventions shape not only what users see but also how they understand their own contributions. Seering et al.~\cite{seering2018socialrolesofbots} describe how bots shape conversational norms in online communities by signaling what is appropriate or valued. Yang et al.~\cite{yang2019socialroleshealthcommunities} find that automated agents affect how users participate and interpret insights in health forums. Geiger et al.~\cite{geiger2017conflictandcooperationbots} demonstrate that automated interventions on Wikipedia can recontextualize user contributions on collaborative platforms. Together, these studies illustrate how system-generated representations can establish frames that shape users' interpretations of their own activities, even when systems are not explicitly designed to interpret personal data.

Similar dynamics may arise when fitness platforms attach AI-generated interpretations to workout records. Platform-authored summaries can foreground specific readings of an activity while downplaying others, potentially narrowing how a workout is understood~\cite{bentvelzen2021reflectionmodel}. These dynamics become salient when system-generated interpretations do not align with an individual's intentions or embodied experience.

\subsection*{Summary}

Prior work shows that making sense of self-tracked data is a situated and embodied process, yet the integration stage, in which data are transformed into interpretations, remains comparatively underexamined. While recent fitness platforms increasingly automate this stage through AI-generated summaries, existing studies focus mainly on AI tools that users voluntarily engage with, such as coaching assistants. Less is known about how people respond when platforms generate interpretations automatically and attach them to activity records by default. Our work helps address this gap by examining individual reactions to Strava's Athlete Intelligence, an AI feature that generates platform-generated workout summaries.

\section{Strava's Athlete Intelligence}

Strava combines individual self-tracking with social interaction. 
The platform presents itself as an inclusive space where ``\textit{everyone belongs on Strava when they are pursuing an active life}''~\cite{strava2024gift}. Users of Strava log workouts, monitor progress, and stay motivated through features such as kudos, comments, and leaderboards, which compare performance on shared routes or segments.

\begin{figure*}[h!]
    \centering
    \begin{subfigure}[b]{0.30\linewidth}
        \includegraphics[width=\linewidth]{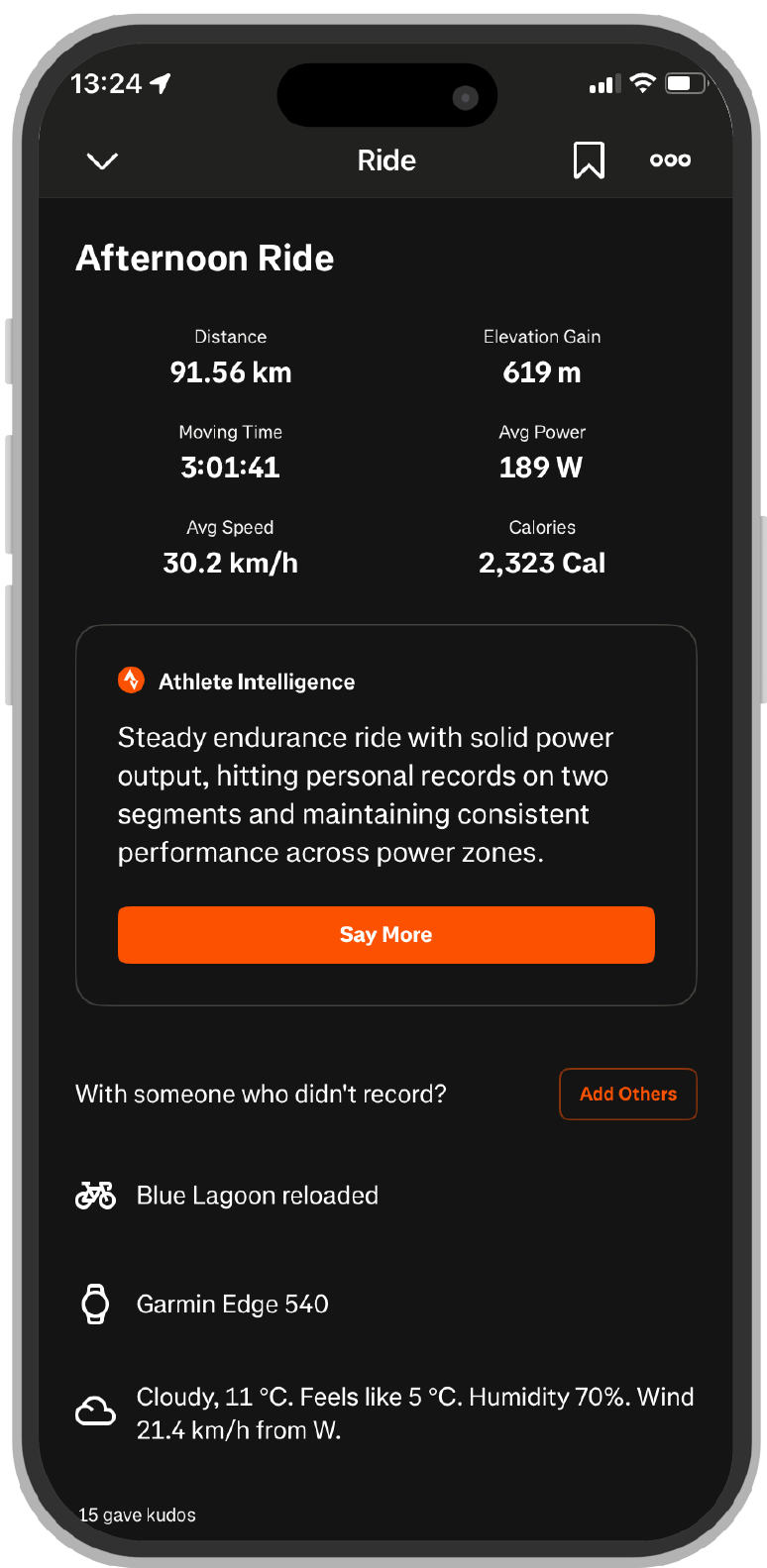}
        \caption{Each tracked activity provides users with an AI summary from Athlete Intelligence (middle of the UI).}
    \end{subfigure}
    \hfill
    \begin{subfigure}[b]{0.30\linewidth}
        \includegraphics[width=\linewidth]{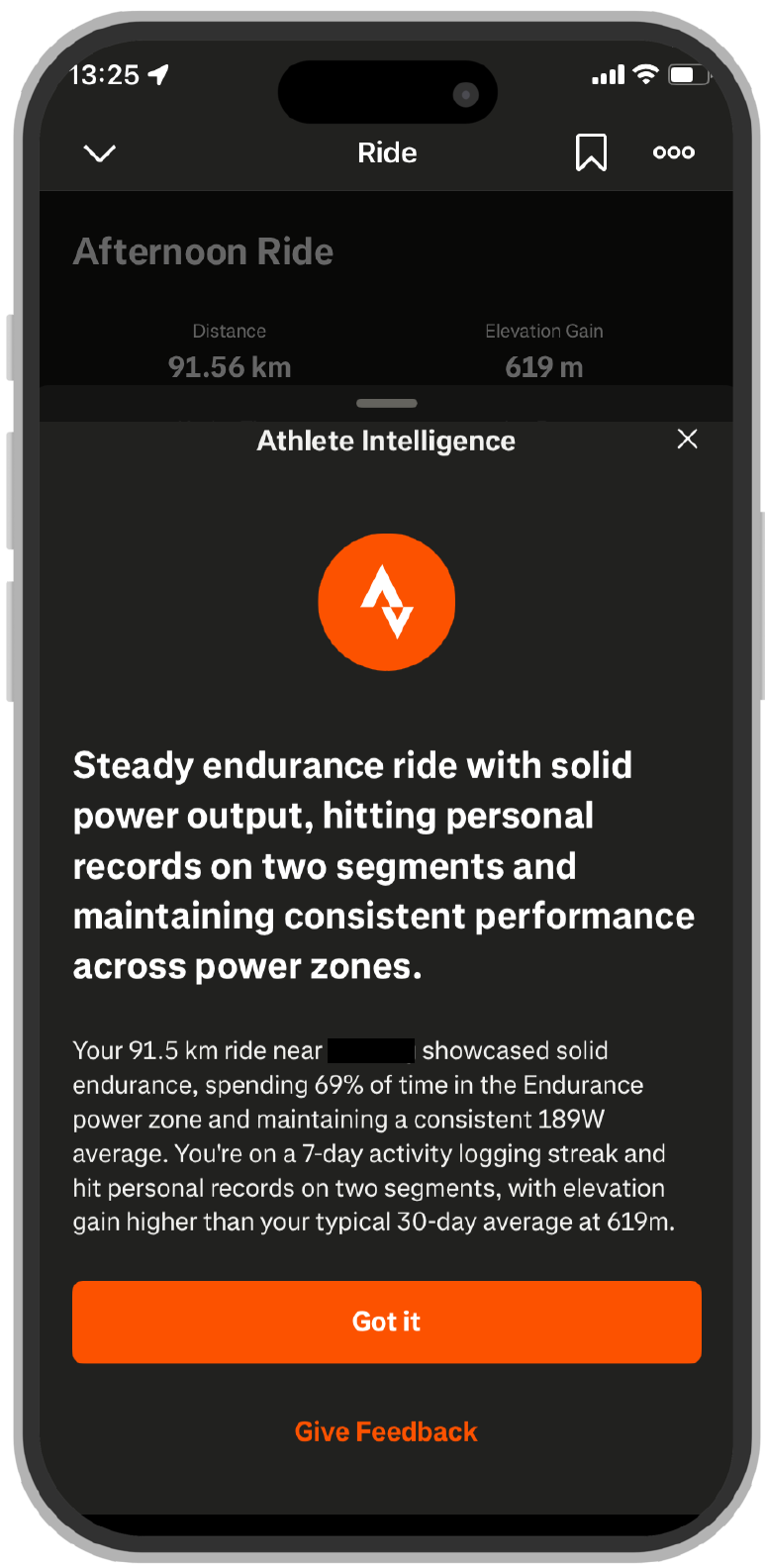}
        \caption{Users can then select [\texttt{Say More}] to expand the summary and receive additional interpretation.}
    \end{subfigure}
    \hfill
    \begin{subfigure}[b]{0.30\linewidth}
        \includegraphics[width=\linewidth]{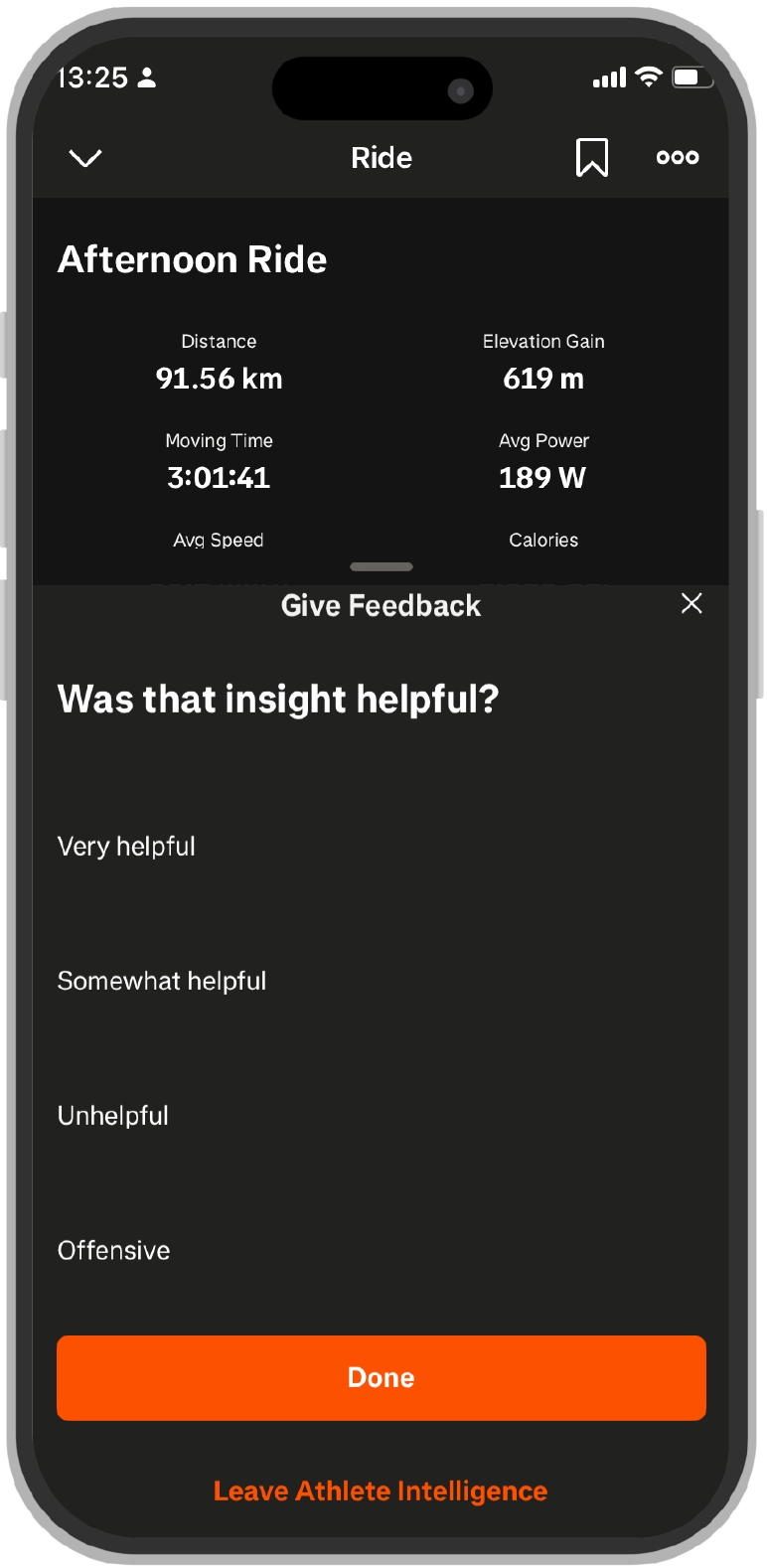}
        \caption{Selecting [\texttt{Give Feedback}] allows users to rate the summary or leave the feature completely.}
    \end{subfigure}
    \caption{Examples of Athlete Intelligence interface elements as seen by Strava users. Screenshots are taken from a member of the research team's personal device.}
    \Description{This figure contains three smaller images showing the Athlete Intelligence interface. The first shows a short AI-generated summary on the post-activity screen. The second shows the ``Say More'' option expanding the summary. The third shows the ``Give Feedback'' option with multiple-choice ratings such as Very Helpful, Unhelpful, Offensive or Leave.}
    \label{fig:Athlete_Intelligence}
\end{figure*}

As part of its effort to enhance how users engage with their data, Strava introduced \textit{Athlete Intelligence} on October 3, 2024, a generative AI-powered feature designed to translate workout metrics into personalized, conversational summaries~\cite{Athlete_Intelligence}. As shown in \autoref{fig:Athlete_Intelligence}, the feature analyzes training trends over the previous month using metrics such as pace, heart rate, power, elevation, and relative effort, and produces short narrative interpretations.
Crucially, these summaries appear automatically within the activity view, positioned below basic metrics such as distance and pace. 
Users encounter the AI's framing as part of viewing their activity, rather than as a separate tool they choose to consult. The generated summary can be expanded through the [\texttt{Say More}] option or evaluated by users through [\texttt{Give Feedback}]. This limits responses to multiple-choice options of [\texttt{Very Helpful}], [\texttt{Somewhat Helpful}], [\texttt{Unhelpful}], [\texttt{Offensive}], or the decision to leave the feature entirely, as seen in~\autoref{fig:Athlete_Intelligence}~(c). 

A long-standing challenge in fitness tracking is that interpreting numerical data requires interpretive work that is unevenly distributed across users~\cite{NarratingFitness,Loerakker2025lived,PhysioLLM}. While some athletes rely on human coaches or their lived training experiences to interpret their data, others struggle to connect metrics to training goals or lived experience. Athlete Intelligence enters this space by performing integration on the user's behalf, translating aggregated metrics into a single narrative account.

\section{Method}

We studied how users reacted to Strava's Athlete Intelligence feature through a qualitative analysis of the \texttt{r/Strava} subreddit\footnote{\url{https://www.reddit.com/r/Strava/}}. Online forums like Reddit provide naturally occurring accounts of technology use and have been used in HCI to examine user concerns, barriers, and responses to new features~\cite{Kou,Norms,Sannon,Gray}.  
We chose \texttt{r/Strava} because it hosts an active community of over 196,000 members who regularly share activities, celebrate progress, and discuss platform features, making it a suitable setting for examining early reactions to Athlete Intelligence.

We analyzed relevant Reddit threads using a qualitative coding approach that combined inductive coding with a structured codebook, allowing us to examine how users responded to AI-generated integrations presented as interpretations of their training.

\subsection{Data Collection}

To identify relevant posts and comments referencing Strava's Athlete Intelligence feature on the \texttt{r/Strava} subreddit, we adopted a keyword-based data collection strategy similar to other studies in contexts such as platform-level surveillance in the gig economy~\cite{Sannon}. We began by generating an initial seed list of 13 terms derived from early observations of the subreddit and language used in Strava's official product descriptions. These initial keywords were:

\begin{quote} 
\texttt{Athlete Intelligence, Strava AI, Strava insights, Strava fitness tracking AI, Strava training recommendations, Strava effort score, Strava fatigue tracking, Strava AI accuracy, Strava recovery score, Strava vs Whoop AI, Strava overtraining detection, Strava AI coaching, Strava machine learning} 
\end{quote}

Our research team collaboratively engaged in iterative discussions to refine this initial list. Each team independently explored the subreddit content and identified candidate keywords from user discussions. This iterative process included assessing the contextual relevance of each term and identifying emerging expressions that aligned with users' descriptions of AI-based fitness technologies. 

To ensure broad coverage and avoid missing relevant conversations that did not explicitly mention ``\textit{Athlete Intelligence}'', we expanded the list to include more general terms related to generative AI and fitness coaching technologies. This approach aligns with established practices in content analysis, where keyword breadth is critical to capturing varied user terminology and framing~\cite{Sannon,RedditBan}. The final keyword set included 24 terms:

\begin{quote} 
\texttt{Athlete Intelligence, Strava AI, Strava insights, Strava fitness tracking AI, Strava training recommendations, Strava effort score, Strava fatigue tracking, Strava AI accuracy, Strava recovery score, Strava vs Whoop AI, Strava overtraining detection, Strava AI coaching, Strava machine learning, AI, artificial intelligence, chatbot, LLM, LLMs, sports analytics, machine learning in sports, AI coaching, AI running assistant, AI-based training, smart coaching AI, athletic data analysis} 
\end{quote}

Using the Python Reddit API Wrapper (PRAW)~\cite{PRAWDocumentation}, we collected all threads and comments from \texttt{r/Strava} containing one or more of these keywords. Our collection window spanned from October 3, 2024 (the public release date of Athlete Intelligence) to March 25, 2025, when we initiated our analysis. This process yielded a dataset of 297 threads and 5,692 comments, authored by 2,611 unique users. 

\subsection{Analysis}

We analyzed Reddit threads discussing Athlete Intelligence using a qualitative, inductive approach. Our analysis combined close reading of posts and comments with iterative coding to surface how users described and negotiated the introduction of Strava's AI-generated summaries as platform-authored interpretations of their activity. To begin, three researchers independently coded the same 50 posts, generating early codes and themes. We then met to compare impressions and refine our focus before proceeding with the full dataset. 

To guide the analysis, one researcher drafted an exploratory framework that organized early reflections into broad analytic categories, each paired with guiding questions (e.g., ``\textit{How do users feel about the nature of Strava AI's feedback, tone, and usefulness?}'' or ``\textit{What technical and functional issues are raised by users?}''). This framework did not serve as a finalized codebook but as a sensitizing structure to support consistency across the team. Guided by this framework, we divided the dataset into three roughly equal parts (with two researchers analyzing 100 posts each, and one analyzing 97). 
In each case, all associated comments were included to capture community reactions. This division was adopted to make the workload manageable while ensuring that every post and its discussion thread received close attention.
 
Our open coding phase generated 76 initial descriptive codes grounded in the data. Through iterative comparison and synthesis, we refined and consolidated these into 53 focused codes, which were then used to generate overarching themes which in our case describe four tensions in users' responses. As part of this process, we also noted how users positioned themselves and others as novices or experienced athletes. These orientations emerged inductively from the data, for example when users described being `new to structured training' or, conversely, emphasized long-term tracking or coaching practices. 
We used these distinctions to examine how interpretive authority was negotiated differently across experience levels.

\subsection{Ethical Considerations}

We collected Reddit data using PRAW, an official Python wrapper for the Reddit API, which complies with Reddit's terms of service, rate limits, and authentication requirements~\cite{PRAWDocumentation}. Data were drawn from the public subreddit \texttt{r/Strava}, where threads and comments are openly accessible without a Reddit account. Because the material was publicly available, our university's research ethics committee determined that this study did not constitute human participant research under Danish guidelines. As a privacy-preserving measure, we report quotes using anonymized identifiers (e.g., ``P-125''). We excluded posts and comments labeled as \texttt{[deleted]}, which indicated users had removed their account or content. Screenshots used in the results were cropped and modified, with AI text paraphrased and numeric values altered, to preserve user privacy while retaining the structure and tone of the original interface.

\section{Results}

This section outlines four tensions in athletes' responses to AI-generated summaries. The tensions are summarized in ~\autoref{tab:tensions} and further reported in detail.

\begin{table*}[h!]
\centering
\caption{Summary of tensions, contributing frictions, and what they reveal about user perceptions.}
\begin{tabularx}{\linewidth}{p{3cm} XX}
\hline
\textbf{Tension} & \textbf{Frictions} & \textbf{What It Reveals} \\
\hline
\addlinespace[6pt]

\textbf{1. AI's Numerical View versus Athletes' Contextual View} 
& AI compares sessions that differ by context (e.g., terrain, weather) or intent; labels intentionally slow activities as underperformance; restates visible metrics without explanation.
& Users expect feedback that explains why numbers differ, not merely that they differ. AI summaries are perceived as shallow, with little consideration of the reasons behind the numbers. \\ \addlinespace[6pt]

\textbf{2. Isolated Summaries versus an Ongoing Training Narrative}
& AI evaluates each workout on its own; produces repetitive or generic comments; overlooks training phases, intent, load, fatigue, and patterns across time.
& Users understand training as cumulative. Feedback that ignores trajectory feels disconnected from how athletes plan and progress. \\ \addlinespace[6pt]

\textbf{3. Fixed AI Tone versus Athlete's Emotional State}
& AI applies uniform positivity regardless of session quality or user mood; misses humour, disappointment, or intentional context; can feel intrusive or emotionally misaligned.
& Users report that the AI misunderstands the emotional tone of their workouts, making its comments seem hollow or insensitive, prompting some to disengage. \\ \addlinespace[6pt]

\textbf{4. A Single AI Voice versus Different Athlete Types} 
& Same feedback for novices and experts; automatic opt-in without control; limited customisation.
& Users differ in what they need from feedback. While some appreciate recognition, many experienced athletes see the feedback as misaligned with their expertise. \\

\hline
\end{tabularx}
\label{tab:tensions}
\end{table*}

\subsection{Tension 1: AI's Numerical View versus Athlete's Contextual View}

In endurance sports, athletes interpret a workout through the context in which it was performed. While an athlete's contextual view can encompass the broader scope of lived experience, such as being tired after work or managing family obligations, in this paper we focus specifically on activity-bound context. Here, we use ``contextual factors'' to refer to elements such as terrain (flat vs. hilly), weather, perceived effort or fatigue, and the purpose of the session, all of which shape how an effort feels and how its outcome is judged.

These contextual factors provide the framework athletes use to determine whether, for example, a run or a ride felt strong, easy, disappointing, or entirely as planned. Strava already collects metadata that reflects many of these contextual factors. However, Athlete Intelligence does not use this information in meaningful ways for the athlete. It evaluates workouts primarily through numerical differences. Athletes interpret their sessions through context; the AI interprets them through numbers. This contrast became clear in examples such as uphill runs being labelled ``slower than usual'', even when the elevation data was available. The AI captured what happened in the workout (athlete ran slower than before) but not why it happened (failing to acknowledge the uphill). When the AI's numerical view did not align with the athlete's contextual understanding of the workout, users experienced friction.

Cyclists and runners described this tension when the AI labelled slower sessions as underperformance, ignoring factors that made those sessions intentionally or unavoidably slower. Mountain bikers noted that technical terrain demands more effort at lower speeds: ``\textit{every ride on my mountain bike is `slower than your weekly average' – yeah, no shit}'' (P-527). Runners reported similar frustrations when hills or weather accounted for pace differences that the AI treated as deficits: ``\textit{It told me my pace was down today... yeah, because I ran up a hill}'' (P-612). Another noted, ``\textit{I had a headwind the entire way out}'' (P-98). In each case, contextual factors shaped how the workout felt but were invisible to the AI's integration.

\begin{figure*}
    \centering
    \includegraphics[width=0.8\linewidth]{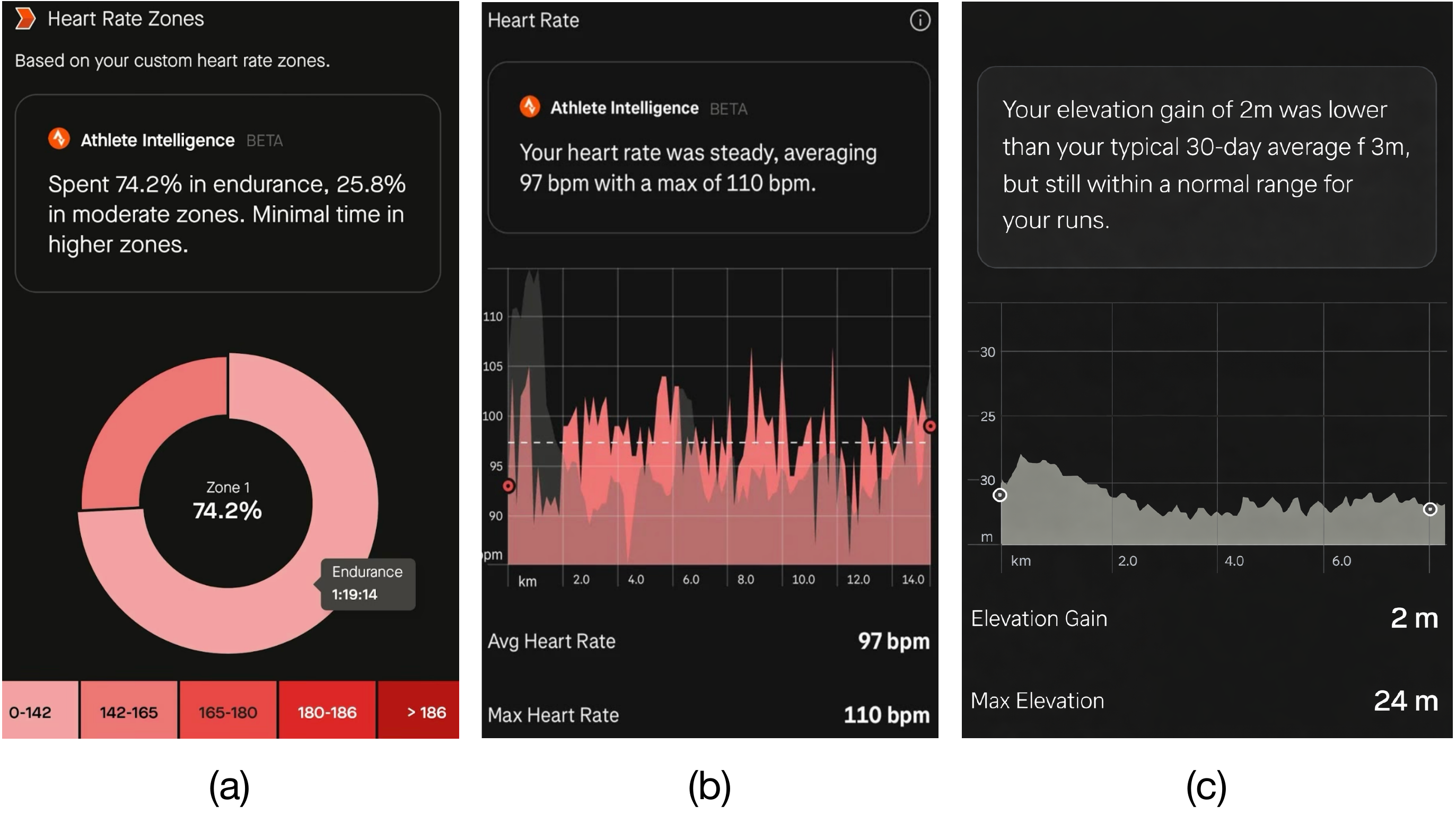}
    \caption{Screenshots of Athlete Intelligence shared on \texttt{r/Strava}, with privacy-preserving adjustments: (a) AI summary restating heart rate zone percentages already shown in the chart; (b) AI describing the shape of the heart rate curve without explaining what caused the pattern; and (c) AI noting elevation is `lower than average' without connecting this to the activities' context.}
    \Description{This figure contains three screenshots shared by users showing redundant AI feedback. The first repeats heart rate zone distributions. The second restates visible trends from a heart rate graph. The third repeats information already shown in an elevation chart.}
    \label{fig:Captain-Obvious}
\end{figure*}

Users also expressed frustration when Athlete Intelligence restated metrics without accounting for the factors that influenced them. Many described the feature as \textit{Captain Obvious}, noting that paraphrasing heart rate zones, pace splits, or elevation patterns did not help them make sense of the effort. As shown in ~\autoref{fig:Captain-Obvious}, users circulated screenshots (modified to preserve privacy) that highlighted how the AI simply repeated visual patterns already visible in the charts. One user compared it to reading a chart aloud: ``\textit{It's like someone pointing at the screen and saying `this line is high' -- thanks, I can see that}'' (P-144). Users wanted the AI to explain the reasons behind the numbers, not just present shallow conclusions or comparisons. Importantly, athletes did not expect the AI to fully reconstruct their lived experience, but to account for basic activity bound contextual factors already associated with the activity when presenting interpretations.

The above examples show that athletes expected AI summaries to acknowledge the contextual factors shaping their efforts. When Athlete Intelligence focused only on numerical differences and ignored these factors, many users found the summaries shallow and incomplete. This disconnect often left athletes frustrated, creating tension between the AI's numerical view and the contextual view on which they relied.


\subsection{Tension 2: Isolated AI Summaries versus Athlete's Ongoing Training Narrative}

Athletes understand their training as unfolding over time. Each workout contributes to a broader arc of building fitness, managing fatigue, and preparing for future goals. Whether they are preparing for a race, following a structured training block, or simply maintaining a routine, athletes interpret each session in relation to what preceded it and what is planned next. Athlete Intelligence, however, treated every activity as a discrete event. Its summaries focused on single-session metrics, with little awareness of patterns or goals beyond a superficial comparison with the previous 30 days.

These frustrations shaped how athletes described the insights they wished the AI would notice. Rather than daily praise or isolated comparisons, they sought recognition of trends across sessions, such as rising fatigue due to changes in training load and the expected timing of recovery. As one user suggested, ``\textit{It could tell you when you're ramping up your training miles too fast}'' (P-249). Another added, ``\textit{Flag when you need a recovery day based on heart rate trends. That'd actually be useful}'' (P-2137). These comments reflected a lack of awareness of patterns spanning days or weeks.

Some athletes compared Athlete Intelligence to tools they already relied on for long-term insight. ``\textit{Garmin tells me if I waited too long between runs. Strava could do this too}'', one user noted (P-176). Others described a desire for basic pattern recognition, positioned between no guidance and the cost of a human coach. As one user put it: ``\textit{There is something that has to exist between zero help and hundreds of dollars for a human coach. AI should help amateur athletes adapt their training to their life conditions}'' (P-550). Even those who appreciated parts of the feature pointed to its potential for identifying longer-term trends: ``\textit{I quite like the four-week evaluation. It helps me see things I wouldn't notice myself}'' (P-20).

The above examples showcase the integration of Athlete Intelligence and contrast how athletes made sense of their training through continuity and expected the AI to at least recognise how sessions connected to one another. However, the AI summaries interpreted each workout using only the numbers from that session, without linking it to any broader training pattern or timeline. As a result, the summaries felt detached from the goals, phases, and progressions that shaped athletes' training over time.


\subsection{Tension 3: Fixed AI Tone versus Athlete's Emotional State}

Athletes often bring strong feelings into their workouts. Pride, frustration, disappointment, relief, or simple fatigue shape their understanding of what the session meant to them. Athlete Intelligence, however, consistently maintained a fixed, upbeat tone that did not adapt to these emotions. This created friction whenever the AI's scripted positivity failed to match how the athlete actually felt.

\begin{figure}[h!]
    \centering
    \begin{subfigure}[b]{0.45\linewidth}
        \includegraphics[width=\linewidth]{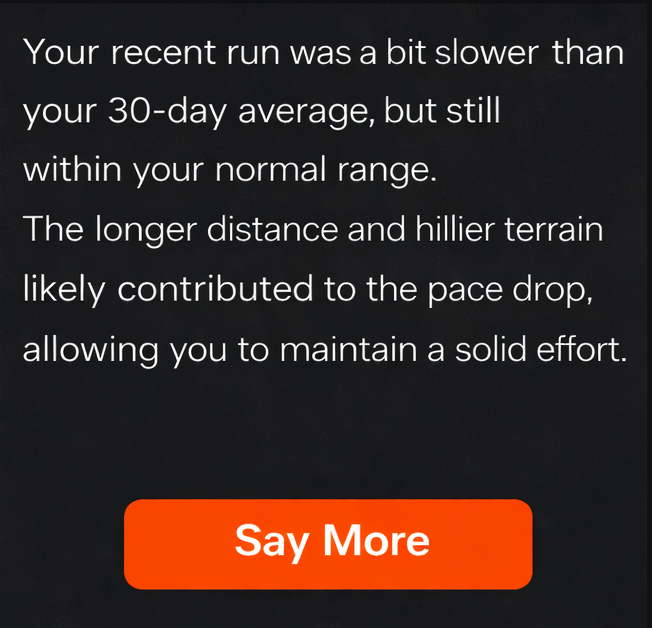}
        \caption{AI framing a successful interval workout as slower than usual pace.}
    \end{subfigure}
    \hspace{2em}
    \begin{subfigure}[b]{0.45\linewidth}
        \includegraphics[width=\linewidth]{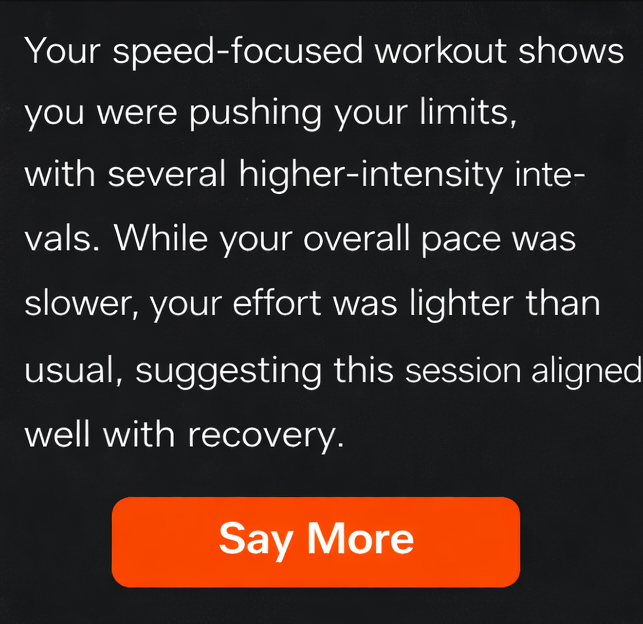}
        \caption{AI ignoring hilly repetition workout and framing it as a lighter effort.}
    \end{subfigure}
    \caption{AI-generated workout summaries illustrating misalignment between automated effort framing and athletes' intended workout experience.}
    \Description{}
    \label{fig:Judgement}
\end{figure}

Some users described the AI's comments as intrusive or out of place. They felt that the system inserted itself into moments they considered personal or reflective, disrupting a space that normally belonged only to them. As one person wrote, ``\textit{I find its comments about my comments super creepy. I don't like that I was opted-in without my say}'' (P-88). For others, the tone did not just feel mismatched, but evaluative: ``\textit{I just want to go for a run without being judged... now one of the few things I actually enjoy ends with a lecture from a bot}'' (P-87). Another user added, ``\textit{Feels like it's looking at my stuff and judging me. I don't want a toxic cheerleader}'' (P-159). These reactions show that the issue was not only the AI's fixed positivity, but the sense that it was occupying a role that athletes had not invited it to take. When the AI's tone entered their personal space with forced encouragement, it felt intrusive against the mood of the workout. ~\autoref{fig:Judgement} provides examples of AI summaries that athletes interpreted as judgmental or oblivious to the personal intentions behind their workouts.

\begin{figure*}
    \centering
    \includegraphics[width=0.8\linewidth]{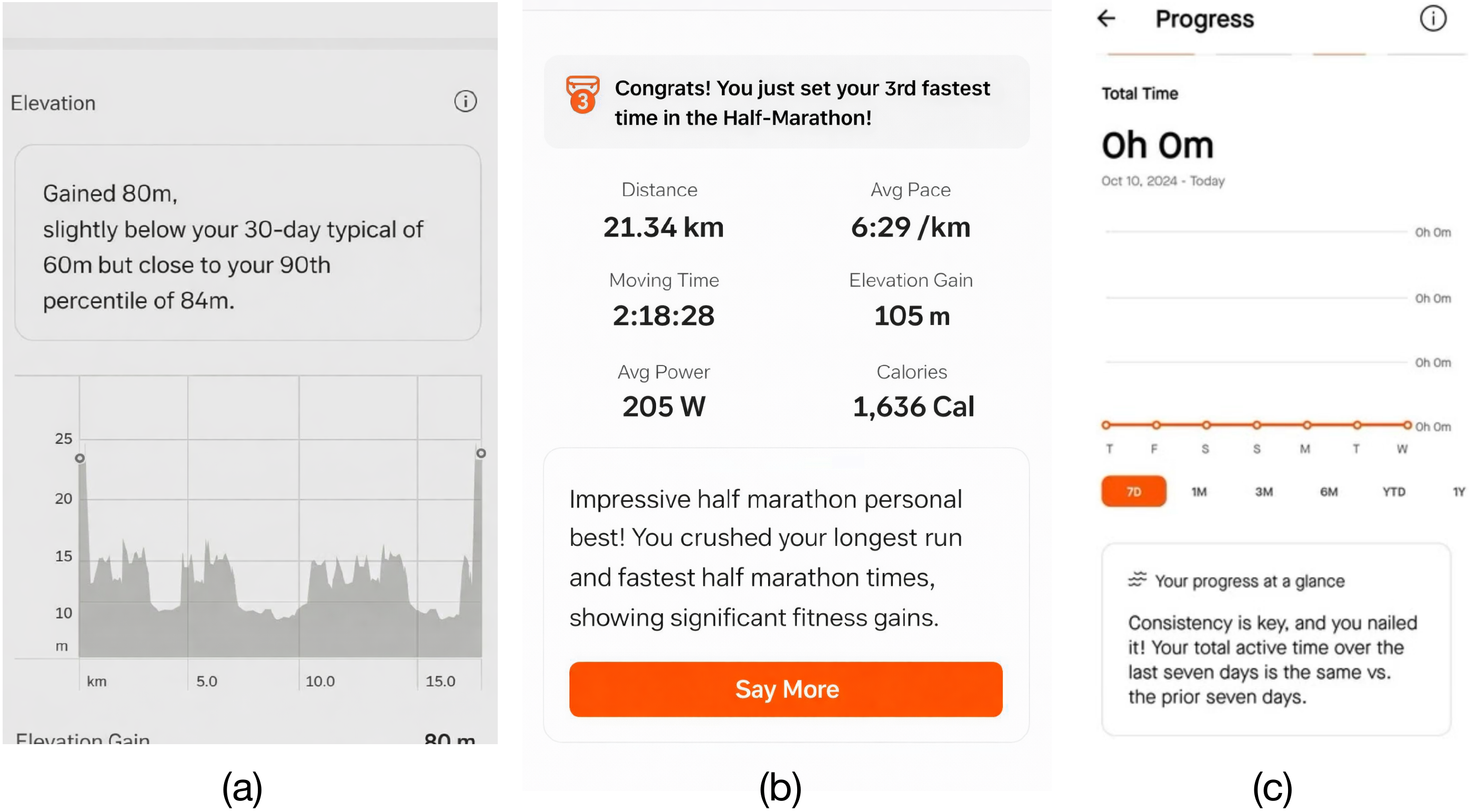}
    \caption{Three screenshots of AI-generated fitness feedback: (a) AI summary restating heart rate zone percentages already shown in the chart, without further explanations; (b) AI describing the shape of the heart rate curve without explaining what caused the pattern; and (c) AI noting elevation is lower than average without connecting this to the activity context.}
    \Description{Three screenshots of AI-generated fitness feedback.}
    \label{fig:Contradictory}
\end{figure*}

Users also observed that Athlete Intelligence defaulted to praise in all situations. What initially felt like encouragement quickly became predictable and empty because the AI summary maintained the same upbeat tone regardless of the workout. As P-35 wrote, ``\textit{Initially I liked the positivity... but then I realised it was always going to be positive, regardless of how bad I am, and so it is meaningless}''. The AI offered identical praise for easy runs, intense workouts, and days when the effort simply did not go well. This mismatch became especially pronounced for athletes experiencing injury, fatigue, or discouraging performances. One user recovering from surgery explained how the AI congratulated them for \textit{crushing} a pace that reflected limited mobility rather than achievement: ``\textit{I can't run due to recent rotator cuff surgery, and AI loved how I crushed this pace. I am slowly returning to running}'' (P-67). Others described how the AI's forced enthusiasm could make disappointment feel even sharper. A marathon runner who narrowly missed a time goal noted, ``\textit{Strava AI really salting the wound with the `just shy' lol}'' (P-27). Instead of acknowledging the difficulty of the moment, the AI responded with lighthearted praise, creating a sharp contrast between the athlete's frustration and the AI's upbeat tone. In these situations, praise felt insensitive because it ignored the emotional weight of falling short of goals and misread the athlete's needs in that moment. 
\autoref{fig:Contradictory} highlights cases where AI summaries contradicted the workout data or applied unearned positivity, reinforcing users' sense that the tone was disconnected from their experience.

Among the Reddit threads further describing this tension, athletes deliberately gave their activities exaggerated or absurd titles simply to expose how poorly the AI handled humour. As one user wrote, ``\textit{I'm gonna try to come up with the most ridiculous titles I can think of, just to see what sort of stupid, meaningless coaching advice it will give me}'' (P-858). Others shared examples of the AI missing obvious jokes or invented scenarios, such as, ``\textit{I wrote `I shit my pants halfway' on a run just to test the AI}'' (P-734), to which the system replied, ``\textit{Despite some issues during your run...}'' Another user joked, ``\textit{Tom Cruise gave me a ride on his motorcycle, and it still told me my pace was a little slow}'' (P-1955). Even everyday mundane sessions revealed the same problem, with P-84 noting, ``\textit{Mine said I `smashed it, mate' after a recovery walk}.''

For some, the constant stream of upbeat praise became irritating rather than encouraging: ``\textit{The only thing Strava AI wants to do is tell you to crush crush crush. I started hating that word so much I turned the AI off}'' (P-242). Others rejected algorithmic encouragement altogether, arguing that praise delivered by code felt fundamentally inappropriate: ``\textit{Any form of encouragement like keep up the good work' that originates in computer code is worse than meaningless, it's insulting}'' (P-1056).

These examples show that when the AI failed to recognize the intended meaning behind a workout, users pushed back by experimenting with exaggerated prompts. Such resistance forms revealed that a fixed emotional script based solely on positivity could not accommodate the range of emotions athletes bring to their training.

\subsection{Tension 4: A Single AI Voice versus Different Athlete Types}

Athletes make sense of their training from their lived training experiences. Experienced or structured athletes often have a well-developed understanding of their data and a clear sense of how different sessions affect their progress. In contrast, novice athletes and those with limited social support on Strava may welcome encouragement. Athlete Intelligence, however, delivers the same evaluative voice to all users, offering identical interpretive styles regardless of athlete type. This creates friction because the feature aligns with the needs of novices and socially isolated users, while often clashing with the expectations of experienced athletes who interpret such feedback as too simplistic to how they see themselves as athletes.

For novices, casual users, or those who received little social feedback; Athlete Intelligence offered a sense of recognition they did not otherwise get from peers on the platform. One user shared, ``\textit{No real humans comment on my activities, bot bro feels nice}'' (P-79). Another described how the encouragement mattered to them during a period of limited social support: ``\textit{This sounds really pathetic but as someone who doesn't have much of a social support network it actually really works for me}'' (P-953). For these users, the AI's generic praise aligned with how they saw themselves: as trying, improving, or simply wanting their effort acknowledged.

For more experienced athletes, the same messages carried very different meanings. They interpreted AI praise not as recognition, but as a misunderstanding of their training knowledge. Many felt the system simply repeated what they already knew. As P-7 wrote, ``\textit{Repeating back to you what you wrote isn't intelligence}.'' Others connected the appreciation for the feature to inexperience. When one user expressed support for Athlete Intelligence, another replied, ``\textit{I'm guessing you're new to structured training?}'' (P-222). Another experienced runner added, ``\textit{Yeah probably because experienced runners like me often already understand their training patterns}'' (P-1566). Across threads, rejecting Athlete Intelligence became a way to signal expertise, whereas valuing it was associated with being a beginner.

This identity tension extended beyond the interpretation of workouts to the rollout of the feature itself. Users who saw themselves as knowledgeable objected to being automatically opted in, reading it as a challenge to their autonomy. As P-1047 put it, ``\textit{I did not opt in and there is no feedback channel. And I can only say if either `this helped' or nothing at all}.'' Others reacted more strongly, treating the feature as evidence that Strava misunderstood what serious athletes wanted. One user wrote, ``\textit{The fact that they waste money on this instead of fixing things that users have asked for for years says everything}'' (P-155). Another linked their frustration directly to their subscription: ``\textit{Was the straw that broke the camel's back for my Premium. Cancelled it after seeing this garbage AI overlay}'' (P-633). Even humour was used to push back: \textit{The AI feature Strava really needs is an AI-powered opt-out button}'' (P-369).

These examples illustrate how Athlete Intelligence intersected with the lived experiences that different athlete types bring to their training. Novices and socially isolated users sometimes welcomed the encouragement, but experienced athletes often read the same messages as patronizing. The tension, therefore, stems less from what the AI said and more from the fact that a single uniform interpretation failed to account for the different athletic types.

\section{Discussion}

Our results identify four tensions based on user reactions to AI-generated summaries of their training data on a large social fitness-tracking platform. In this section, we examine how these tensions arise from differences between how AI-generated summaries interpret workouts and how athletes make sense of their training in practice. We then discuss their implications for the design of AI-integrated self-tracking systems and consider their relevance for longer-term design decisions.

\subsection{How AI-Integrated Feedback Shapes Interpretation in Fitness Tracking}

Early models of self-tracking describe reflection as a stage that follows data collection, in which people review and reflect on their data in relation to goals and routines~\cite{li2010personalinformatics}. Later research highlights that reflection is rarely linear~\cite{niess2018trackergoal,Loerakker2025lived}. Instead, people make sense of their data over time, shaped by bodily sensations~\cite{epstein2020exploring}, changing goals~\cite{epstein2015livedinformatics}, and everyday life~\cite{bentvelzen2022revisitingreflection,rooksby2014livedinformatics}. More recent work suggests that AI systems can support reflection by organizing or summarizing data, while interpretation itself remains a user-driven process~\cite{NarratingFitness,GiveandTake}.

Our findings highlight that the automatic summarization and integration of fitness data does not always align with how users interpret their workouts in practice. In the case of Athlete Intelligence, AI-generated feedback introduces its own interpretation of the workout, framing the activity alongside athletes' own emerging interpretations. User tensions with AI were not primarily driven by incorrect feedback, but by how the feedback framed their activity. Athletes often described the summaries as judgmental, especially when intentionally slow workouts were reduced to short-term performance comparisons rather than situated within a more personal training timeline, such as preparation for an upcoming goal. This framing left limited room for athletes to interpret their effort in ways that aligned with their own understanding of training.

We identify two design characteristics that amplified these tensions. First, fixed comparisons, such as 30-day averages, treated individual sessions in isolation from broader training trajectories or intentions. Such performance-oriented framings can reinforce normative ideals while failing to support different athlete types and goals~\cite{Schneider,Fitter,Ortega,Rosa}. In our findings, this contributed to feedback that framed workouts according to standardized expectations rather than athletes' situated training intentions.

Second, a stable, upbeat tone applied a uniform evaluative framing across diverse situations. This reflects Eikey et al.'s description of feedback that flattens emotional nuance~\cite{eikey2021beyondreflection}. Similar patterns have been observed in health tracking contexts, where affect-blind feedback can feel dismissive or manipulative~\cite{alberts2024computers,sun2024can}. Skilton and Cardinal~\cite{ToxicPositivity} describe a related tendency in generative AI, in which ``toxic positivity'' emerges as a safe default that avoids critique while distorting user intent.

For design researchers, these findings underscore that AI-integrated feedback is not neutral. Feedback actively shapes how workouts are understood by foregrounding certain aspects of performance and ignoring others. On social fitness platforms, users already make sense of their training through experience and interaction with others, with AI introducing an additional layer of interpretation alongside these existing reflective practices. The usefulness of AI-integrated feedback depends not only on how clearly it explains the data but also on whether the design allows users to retain control over how that explanation is interpreted. 

Our findings also suggest that AI is not equally suited to all aspects of self-tracking. While athletes are well equipped to interpret individual workouts through contextual understanding and embodied experience, AI may be more useful in identifying patterns across sessions, such as changes in training load, recovery, or longer-term trends. This highlights a complementary role for AI in supporting pattern detection across sessions, while leaving the interpretation of individual workouts to athletes.

\subsection{Designing AI-First Integration in Self-Tracking Systems}

The tensions identified in our findings point to a mismatch between how AI-generated feedback presents workout summaries and how athletes interpret their training in practice. AI-generated feedback often treats integration as the production of a single, consolidated account of an activity. Athletes, by contrast, understand their training as an ongoing and open-ended process shaped by context, experience, and evolving goals~\cite{rapp2020selftrackingsport,WOODS2019318}. This perspective is reinforced by Tholander et al.~\cite{Tholander}, who demonstrates that athletes often rely on bodily sensations as a primary source of knowledge when interpreting performance, rather than treating metrics as definitive representations.

This mismatch reflects prior work in personal informatics, which shows that automated analyses can offer limited value in domains where users already possess interpretive expertise~\cite{rapp2023livedexptechnologies,Moore}. When integration is embedded at the platform level and delivered automatically, it becomes difficult to disengage from its framing. Similar dynamics have been observed in other community contexts, such as Wikipedia~\cite{geiger2017conflictandcooperationbots} and health forums~\cite{yang2019socialroleshealthcommunities}, where automated contributions shape how information is interpreted.

Personal informatics research describes self-tracking as a process involving preparation, data collection, integration, reflection, and action~\cite{li2010personalinformatics,epstein2015livedinformatics}. While these stages are not strictly linear, they provide a useful lens for understanding how AI-generated summaries intervene in ongoing sense-making practices. Rather than extending integration to produce complete interpretations, AI support in self-tracking contexts should be designed to align with how users already interpret their data. Based on this, we outline design implications for AI-first integration in self-tracking systems.

Current AI-generated feedback is implemented in a rigid and conclusive manner, often presenting workout summaries as finished interpretations. This conflicts with PI models, which frame the integration stage as preparing data for reflection rather than completing reflection on the user's behalf~\cite{epstein2015livedinformatics,epstein2020mappingpersonalinformatics}. \textbf{We therefore recommend designing AI-generated feedback as a mediator that narrates observations to support users' reflective processes}. Such narration can draw attention to patterns or facts in the data that might otherwise be overlooked. Prior work in fitness tracking~\cite{NarratingFitness} shows that narrative text is a powerful entry point for different user types because of its seamless readability. However, this entry point should not be used to produce surface-level conclusions that risk inducing rumination~\cite{eikey2021beyondreflection,Niess_Rumination} rather than reflection. \textbf{We further recommend avoiding overly positive narrative feedback, as it can lead to abstract, context-poor integration which can intensify negative reactions when users attempt to correct or clarify the AI's feedback}.

PI research shows that self-tracking is rarely continuous and often includes lapses and periods of non-use~\cite{niess2018trackergoal}. Our findings, in contrast, show that AI-generated feedback commonly assumes continuous tracking and a default readiness for reflection. However, in fitness tracking, pauses are often due to fatigue, injury, or life transitions rather than disengagement~\cite{lapses}. Treating such moments as failures risks producing narrative AI-feedback as incomplete to users. \textbf{We therefore recommend designing AI feedback to recognize pauses, missing data, and irregular use as valid states when integrating data}. Additionally, allowing users to choose their own temporal ranges for reflection, rather than relying on fixed comparison windows, can better reflect athletic lifestyles that include phases of peak training, recovery, and off-season periods~\cite{TrainingLoad,Coşkun21072023}.

AI-generated feedback currently focuses narrowly on sensor data during integration, while overlooking textual annotations that capture rich qualitative context. Prior work shows that athletes often rely on such annotations to express intent and make sense of workouts in ways that metrics alone cannot capture~\cite{niess2018trackergoal,bentvelzen2022revisitingreflection}. Our findings showed that textual annotations contained important information about conditions, intent, and atypical events (e.g., bad sensor readings). Yet, this information was treated as peripheral or ignored entirely. As a result, current AI-generated feedback relies on an extrinsic, goal-oriented mode of integration that does not reflect how all athletes engage with their data. PI research shows that self-tracking also supports curiosity and self-exploration, not only goal pursuit~\cite{Moore,choe2017reflectpersonaldata}. Consistent with this broader orientation, athletes in our study (see tension 3) often attempted to engage more actively with the AI, including through dialogical interaction, despite it not being designed for dialogue. \textbf{We therefore recommend extending AI support into the collection stage of the PI model by enabling dialogical input that captures rich qualitative context}. This design direction can support more playful and exploratory engagement with data~\cite{SelfCare} by allowing athletes to reflect and build self-knowledge on their own terms, rather than being pushed toward a fixed narrative~\cite{rapp2017knowthyself}.

\subsection{The Long-Term Usefulness of AI-Generated Feedback in Social Fitness Platforms}

Beyond understanding how AI-generated feedback shapes reflection, our findings raise questions about its long-term usefulness in social fitness platforms. Prior work in fitness technologies~\cite{GiveandTake,GPTCoach,MoveAI} suggests that the long-term value for users depends on whether systems continue to support users' sense-making as their training practices evolve. Static AI feedback that does not adapt to changing goals or experience levels can become less relevant over time. The tensions were evident when experienced athletes opted out of the feature almost immediately upon encountering the AI feedback. 

We raise the question of whether AI-mediated integration is necessary in platforms where users already engage in established interpretive practices. On social fitness platforms, these practices can be achieved through activity annotations and peer interactions. Our findings, along with prior work~\cite{bentvelzen2023fitnesstrackers,RiversStrava}, show that these practices are well-accepted and also develop through users' continued participation~\cite{Couture02012021}. Here, AI-generated feedback in its current form does not address a missing capability within these practices. Instead, it adds an additional layer of interpretation that creates friction among established practices. 
Strava presented AI-generated feedback as a way to support novice users as they transition towards becoming experienced users~\cite{strava2024gift}. 
However, our findings suggest that the AI-generated feedback, in its current static form, offers only short-term value and risks losing utility once the novelty wears off.

\subsection{Limitations}

Our analysis of Reddit forums gave access to authentic discussions of a new AI feedback feature, but this approach also has limitations. Our analysis focused on self-trackers in the Strava subreddit \texttt{r/Strava}, excluding users who do not participate in this subreddit. This narrows the diversity of perspectives represented. We also acknowledge that Strava offers many self-tracking features beyond Athlete Intelligence. To the best of our knowledge, however, Athlete Intelligence is the first explicitly branded AI feature on the platform, making it a distinct focus for this study. At the time of data collection, the feature was only available to paying subscribers. According to Strava, subscribers are more likely to be early adopters of new technologies and tend to be more serious athletes with higher expectations for data-driven reflection~\cite{StravaUserProfile}. Reddit discussions may over-represent critical perspectives, as such forums often serve as spaces for expressing dissatisfaction. Our findings therefore capture tensions articulated by users who found the feature misaligned, rather than representing all Strava users. Finally, although we applied a wide set of keywords, some relevant posts may have been missed, and our dataset may not capture the full spectrum of perspectives.

\section{Conclusion}

This paper examines how Strava users reacted to Athlete Intelligence, an AI feedback feature that summarizes workouts. Drawing on Reddit discussions, we find that while the feedback sometimes simplified interpretation, it often presented interpretations that did not align with how athletes understood their training, repeated obvious information, or conflicted with existing ways of making sense of activity data. Our findings contribute to HCI and design research by showing how AI-generated feedback introduces an additional interpretive layer when the platform, rather than the user, leads the integration of data. We identify tensions that arise when generic summaries do not align with the personal and social ways athletes make sense of their training, and highlight the need for feedback that supports, rather than replaces, user interpretation. As AI becomes more prevalent on fitness platforms, the challenge is to reduce entry barriers for newcomers without undermining established practices. Our findings suggest that AI is better suited to identifying patterns across sessions, such as trends in training load or recovery, while the interpretation of individual workouts remains grounded in athletes' contextual view. Designing feedback as an optional, adaptable layer rather than a fixed summary can help AI support sense-making while maintaining user control.

\begin{acks}
We would like to thank our reviewers for their positive reception of the work, and their helpful and constructive feedback to improve the paper.
\end{acks}

\bibliographystyle{ACM-Reference-Format}
\bibliography{references}

\end{document}